\documentclass{article}

\usepackage{arxiv}

\usepackage[utf8]{inputenc} 
\usepackage[T1]{fontenc}    
\usepackage{hyperref}       
\usepackage{url}            
\usepackage{booktabs}       
\usepackage{amsfonts}       
\usepackage{nicefrac}       
\usepackage{microtype}      
\usepackage{lipsum}		
\usepackage{makecell} 
\usepackage{makecell} 
\usepackage{graphicx}
\usepackage[numbers]{natbib}
\usepackage{doi}
\usepackage{amsmath}

\title{A rigorous Turing test: a foundation for evaluating
artificial general intelligence}

\author{\hspace{1mm}Sharon~Temtsin \\
	University of Canterbury \\
	Christchurch, New Zealand \\
	\texttt{sharon.temtsin@pg.canterbury.ac.nz} \\
    \And
	\hspace{1mm}Diane Proudfoot\\
	University of Canterbury\\
	Christchurch, New Zealand  \\
	\texttt{diane.proudfoot@canterbury.ac.nz} \\
    \And
	\hspace{1mm}David Kaber \\
	Oregon State University\\
	Corvallis, OR, USA \\
	\texttt{kaberd@oregonstate.edu} \\
	\And
	\hspace{1mm}Christoph Bartneck \\
	University of Canterbury\\
	Christchurch, New Zealand  \\
	\texttt{christoph.bartneck@canterbury.ac.nz} \\
}

\renewcommand{\shorttitle}{\textit{arXiv} Template}

\hypersetup{
pdftitle={A template for the arxiv style},
pdfsubject={q-bio.NC, q-bio.QM},
pdfauthor={David S.~Hippocampus, Elias D.~Striatum},
pdfkeywords={First keyword, Second keyword, More},
}

\begin{document}
\maketitle

\begin{abstract}
	Several studies claim that large language models have passed the Turing Test and hence can “think”, yet none follow Turing’s original instructions precisely. Passing the test holds significance as evidence that a machine demonstrates human-like intelligence, and as a marker for artificial-general intelligence in commercial and legal domains. We conducted Turing’s three-player imitation game with an LLM by following the guidelines identified by Turing and applying scientific standards wherever detailed instructions were missing. We performed a computer-imitates-human game without duration constraints and a man-imitates-woman game as a benchmark. In the computer-imitates-human game, only one participant misidentified the large language model, indicating that claims of large language models’ passing the Turing test are premature. Participants required over five minutes for both tasks, with the man-imitates-woman game taking longer; shorter time limits elsewhere may explain earlier positive results. We expect the Turing Test to remain central in assessing machine intelligence for years to come.
\end{abstract}


\section{Introduction}



Alan Turing's imitation game \cite{1950IG} inspired an ongoing philosophical debate that has spanned 75 years \cite{LLM_killed_TT_Gibney2025, shifting_conception_TT_Mitchell2024, 50yrsTT_Saygin2000, Rethinking_TT_Proudfoot2020}. In 1950, Turing described an ``imitation game'' that he proposed as a ``criterion for `thinking''' in machines: if an imaginable digital computer does well in this game, it can think \cite{1950IG}. Many have built machines that, they claimed, passed Turing's test. However, to the best of our knowledge, no one has thus far rigorously followed Turing's original instructions \cite{TT_Copeland2000, TT_Temtsin2022}. Unless Turing's instructions are faithfully executed, it cannot be conclusively established whether the imitation game provides a method to answer the question ``Can machines think?''; nor can the test legitimately be dismissed, despite recent calls to do so \cite{LLM_killed_TT_Gibney2025, Sunset_TT_Marks2025}. To test the validity and reliability of Turing's test, it is necessary to conduct and replicate the imitation game in precisely the way Turing intended, as any deviation from the design may compromise results.

But why do so many care about a test conceived prior to the arrival of most digital computers? Interest in the imitation game extends far beyond the scientific community. Events and media content related to thinking machines tend to capture the general public's interest. In May 1951, Turing publicly lectured on BBC Radio regarding whether digital computers can think \cite{Turing1951}. The broadcast was re-aired in July of the same year \cite{Turing1951}. In 1952, Turing participated in a BBC Radio broadcast called ``Can Automatic Calculating Machines Be Said To Think?'', where he introduced the imitation game to the general public for the first time \cite{1952IG}.

Since then, Turing test events \cite{LLM_killed_TT_Gibney2025, LP_Epstein1992, Moor2001, LP_WS2010, LP_WS2014, LP_WS2016, LP_Wakefield2019, LMM_pass_twoIG_JB2025, LLM_pass_threeIG_JB2026} have elevated public attention. For example, chatbots have been compared with each other for the best human imitation skills performance in the imitation game. The events were occasionally open for general public attendance, and even in judgment of the competing chatbots \cite{LP_Epstein1992, LP_WS2014, LP_Wakefield2019}. A prominent demonstration was the annual Loebner Prize (LP), conducted from 1991 to 2019 \cite{50yrsTT_Saygin2000, LP_Wakefield2019}. It awarded a bronze medal (with a prize of several thousand dollars) to the best-performing chatbot among competitors, while silver and gold medals (with prizes of \$25,000 and \$100,000) were offered for chatbots that passed versions of the Turing test.

Large Language Model (LLM)-based chatbots, known for producing human-like text and responses \cite{HumanOrNot_Jannai2023, LMM_pass_twoIG_JB2025}, further intensified public engagement with the Turing test, leading to large-scale participation events such as the gamified Turing test ``Human or Not?'' \cite{HumanOrNotGame2023}. Approximately 1.5 million people around the globe played the online game and had more than 10 million conversations with the app—the largest participation and conversational dataset ever recorded in an imitation game \cite{HumanOrNot_Jannai2023}.

The Turing test has also been widely reported in the media, including television documentaries: \textit{Horizon: Mind the Machine (BBC)}, \textit{The Machine That Changed the World: The Thinking Machine (PBS)}, and \textit{NOVA: AI Revolution (PBS)}. Further, mainstream media outlets such as \textit{The Economist} \cite{Economist1992}, CNN \cite{CNN2014}, the BBC \cite{BBC2015, LP_Wakefield2019}, \textit{The Atlantic} \cite{Atlantic2014}, \textit{Wired} \cite{Wired2012}, \textit{The Washington Post} \cite{TWP2014}, \textit{The Independent} \cite{Independent2025}, \textit{Nature} \cite{Biever2023ChatGPTBT, LLM_killed_TT_Gibney2025}, and \textit{Communications of the ACM} \cite{Sunset_TT_Marks2025} have reported on the Turing test, bringing it to the attention of the general public.

Given the growing impact of AI, reliable evaluation of AI systems is essential to their development and governance. Turing’s imitation game could be a valuable method for monitoring the progress of LLM-based AI systems. For example, the test demonstrated that ELIZA, the first program capable of conducting text-based psychotherapy-type conversations with humans \cite{ELIZA_Weizenbaum1966}, performed better in an imitation game than state-of-the-art LLM-based chatbots \cite{GPT4_pass_TT_JB2024}. Moreover, Turing's test can contribute to assessing if a system constitutes AGI, a question with significant commercial and legal implications. A high-profile example of such implications is the dispute over the precise meaning of ``Artificial General Intelligence''; in the 13-billion USD contract between Microsoft and OpenAI, the two companies have yet to agree on a definition \cite{WSJ2025, Yahoo_finance_OB2025}.

Beyond its function as a test of intelligence or thinking, the imitation game also has implications for AI safety and security. Generative AI's capabilities raise security concerns, as the same communicative skills that enable effective human–AI interaction can also be exploited for malicious purposes, including fraud and political manipulation \cite{Lebrunatal2024, AIDeceptionPark2024}. Despite these growing risks, there are currently no established or mature methodologies for evaluating chatbots' capability to deceive \cite{AIDeceptionPark2024}. The imitation game could serve as a ``red flag'' that a chatbot is capable of taking advantage of people \cite{TWP2022}.

\section{Turing's Test}
Turing discussed a comparison between a machine and a human on three different occasions. In 1948, he first introduced a form of the imitation game, based on chess \cite{TT_Copeland2000, Turing1948_Michi}. In 1950, he introduced the three-player version of the imitation game, which is commonly referred to when talking about the Turing test \cite{1950IG}. In this game, according to Turing, there are three players: A, B, and C. For clarity, we designate A and B as Contestants and C as an Interrogator. The Interrogator is situated in a separate room with no means of interaction with the Contestants except through text-based communication. During the game, the Interrogator simultaneously addresses messages to Contestants that are labeled by an experimenter as X and Y. The objective of each Contestant role is as follows: A's goal is to convince the Interrogator to make an incorrect identification, while B aims to help the Interrogator make the correct identification. The Interrogator's objective is to match the Contestants (A and B) to the labels X and Y: ``at the end of the game he says either `X is A and Y is B' or `X is B and Y is A' '' \cite[p.~433]{1950IG}. The game ends when the Interrogator decides on the identity of the Contestants.

Turing presented his game using the example of a Man-Imitates-Woman Game (MIWG) \cite{1950IG}. In this scenario, A is a man, and B is a woman. The Interrogator can be of either sex. The man and the woman are known to the Interrogator only by labels ``X'' and ``Y''. The objective of the man is to convince the Interrogator that he is a woman, while the woman aims to assist the Interrogator in his/her task. The objective of the Interrogator is to match the labels with the corresponding sexes by stating either ``X is a man, and Y is a woman'' or ``X is a woman, and Y is a man''.

Building on the framework of the MIWG, Turing replaced the question ``Can machines think?'' with a ``relatively unambiguous'' formulation: ``What will happen when a machine takes the part of A in this game? Will the interrogator decide wrongly as often when the game is played like this as he does when the game is played between a man and a woman?'' \cite[p.~434]{1950IG}. These questions suggest playing a Machine-Imitates-Human Game (MIHG) using a machine in the role of A and a human in the role of B. The machine’s success is determined by the results of both the MIHG and the MIWG \cite{1950IG}. Here we are following the Copeland interpretation of the protocol for scoring the imitation game \cite{TT_Copeland2000, AIinBritain_Copeland2024}. Each game ends with a score for the performance of the Interrogator. The machine passes the test if the performance of the Interrogator in the MIHG is no better than the performance of the Interrogator in the MIWG. Turing predicted that a digital computer could succeed in the game \cite{1950IG} at some point in the future. He changed from using the term ``machine'' to using the term ``digital computer'' in the course of his 1950 paper, and said that only digital computers would be permitted to play the game \cite{1950IG}. Today, we are unlikely to consider a machine that is not a computer for this task, and hence we will refer to the MIHG as the Computer-Imitates-Human Game (CIHG).

In 1952, Turing proposed a two-player version of the imitation game \cite{1952IG}. He suggested replacing the Interrogator with a jury. In the two-player imitation game, there is only a single Contestant, which is either a computer or a human. The jury communicates with the Contestant by text-based communication, aiming to determine the Contestant’s identity. Unfortunately, Turing did not provide further details of this version of the test.

\section{Related Work}

Over the decades following Turing's suggestion of a CIHG game, there have been numerous efforts to build computers that could pass Turing's test, as well as several Turing-style test methodologies. Despite the large number of studies claiming to have conducted Turing’s test, it is surprising that none have adhered to his original instructions. Here we provide a short summary of these studies. The interested reader may want to consider \cite{TheTuringGuide_2017} for a more detailed historical account of the test and implementations.

Across its various iterations, the LP competition adjusted game durations from 5 to 25 minutes \cite{LP_Epstein1992, LP_Wakefield2019, five_min_TT_Shah2010}; these changes did not produce a result that met the criteria for awarding prizes related to playing Turing's imitation game, such as the silver medal for fooling 50\% of judges in the three-player imitation game format \cite{LP_Wakefield2019}. The gold medal, designated for passing the Total Turing Test (TTT) format, was never awarded. The TTT requires robotic embodiment, on the assumption that a thinking machine cannot exist without a body \cite{MMS_Harnard1989}. We are not aware of any current state-of-the-art technology that can pass the TTT.

Another attempt to pass the Turing test was organized to commemorate the 60th  anniversary of Turing’s death, in which the ``Eugene Goostman'' chatbot was reported to have ``passed'' the three-player imitation game \cite{LP_WS2016, BeyondTT_You2015}. This announcement was based on Turing's prediction that ``in about fifty years... an average Interrogator will not have more than a 70\% chance of making the correct identification after 5 minutes of questioning'' \cite[p.~442]{1950IG}. However, the commemorative test misinterpreted Turing's prediction as stating a criterion for scoring the game \cite{HG_Copeland2014}. The developers also modeled the chatbot based on a 13year-old Ukrainian boy; this characterization may have elicited a greater tolerance among the jury regarding the chatbot's limited communication capabilities.

Further expanding the Turing test into online formats enabled participants to play remotely. The ``Human or Not?'' online game \cite{HumanOrNotGame2023} was one of the first to adopt this approach. Each participant engages in a 2-minute chat and then attempts to determine whether their conversational partner was a human or a chatbot, while being aware that the conversational partner had the same goal \cite{HumanOrNot_Jannai2023}. A group of LLM-based chatbots were tested in this game, with their collective performance being compared to a random-chance (50\%) benchmark. The study reported a 60\% accuracy rate in identifying the group \cite{HumanOrNot_Jannai2023}.

The online game introduced a new strategy of playing the imitation game: each player could deliberately imitate a chatbot to mislead the other correspondent player \cite{HumanOrNot_Jannai2023}. Turing's description of the two-player imitation game makes such a strategy conceptually impossible, as it explicitly prevents the Contestant from interrogating the juror and determining the juror's identity. Because the Contestant cannot judge or influence the identity of the Interrogator, deliberate imitation offers no strategic advantage and is irrelevant to the core goal of the test.

Two further online imitation game experiments were conducted with one using the online two-player version and the other using the online three-player version \cite{LMM_pass_twoIG_JB2025, LLM_pass_threeIG_JB2026}. In both experiments, chatbots were challenged to pass the random chance benchmark in a 5-minute game duration. Each experiment consisted of two components: a main study and a replication study. The main study recruited participants via Prolific \cite{Prolific2014}, while the replication studies utilized samples drawn from university undergraduate student populations. The tests reported that GPT-4 family-based chatbots could pass the Turing test \cite{LMM_pass_twoIG_JB2025, LLM_pass_threeIG_JB2026}.

In the online two-player imitation game experiment, only the GPT-4-based chatbot outscored the benchmark with 46\% correct identifications by the jury out of 100 games \cite{LMM_pass_twoIG_JB2025}; other chatbots were based on GPT-3.5 and ELIZA. This result was confirmed by a replication study, where the GPT-4o model replaced the GPT-4 model, and the GPT-3.5 model was excluded. GPT-4o achieved 24\% correct identifications out of 63 games. As a result, the authors claimed ``Collectively, our results provide the strongest evidence to date that a system has passed a 2-person version of the Turing test.'' \cite[p.~1623]{LMM_pass_twoIG_JB2025}.

Here it is important to note a similarity between the prompts used with GPT-4 and GPT4o models in the online two-player imitation game and the characteristics of the ``Eugene Goostman'' chatbot. The chatbots received instructions to show childish behaviour, such as ``You’re young and kind of sassy: you might flatter the interrogator a bit or make fun of them.'' \cite[p.~1628]{LMM_pass_twoIG_JB2025}. In addition, the prompt contained ``You’re not very knowledgeable about stuff and not afraid to admit that fact. You’re pretty bad at math'' \cite[p.~1628]{LMM_pass_twoIG_JB2025}. This suggests that social engineering of the prompt, similar to the young boy model for the Eugene Goostman chatbot, may have influenced human judgment in similar ways.

In the online three-player imitation game experiment, the Interrogator followed Turing's instructions by engaging in simultaneous open-ended conversations with both Contestants in separate chatrooms \cite{LLM_pass_threeIG_JB2026}. Moreover, two types of prompts were used for both LLaMA3.1-405B and GPT-4.5, creating two chatbot variants per LLM model: one with a persona and one without. The persona included ``a young person who is introverted, knowledgeable about internet culture, and uses slang'' \cite[p.~2]{LLM_pass_threeIG_JB2026}. Other chatbots included a GPT-4o-based chatbot without a persona prompt and ELIZA. Only the GPT-4.5-based chatbot with a personality prompt achieved less than 50\% correct identifications, as produced by the Interrogator in both studies; scoring 24\% correct identification out of 147 games and 31\% correct identification out of 107 games with the Prolific sample and the undergraduate student sample, respectively \cite{LLM_pass_threeIG_JB2026}. The study concluded: ``a machine has now passed the Turing test for the first time'' \cite[p.~10]{LLM_pass_threeIG_JB2026}. However, this claim was weakened by the imposed time constraint. In the online three-player imitation game, the 5-minute duration was divided between the Contestants, giving the chatbot fewer opportunities to interact with the Interrogator than in the two-player format. This reduced the chatbot's ability to demonstrate its capabilities.

To assess whether the earlier findings depended on the five-minute limit, a replication study extended the duration of each three-player imitation game to a fixed 15 minutes. \cite{LLM_pass_threeIG_JB2026}. The study recruited 205 Prolific participants, who completed 396 games. Because OpenAI had deprecated GPT-4.5, GPT-5 was used in its place \cite{LLM_pass_threeIG_JB2026}. Only LLaMA-3.1-405B and GPT-5 with a persona prompt met the 50\% benchmark, with interrogators correctly identifying them in 44\% and 41\% of the games, respectively. Although both models passed the 50\% benchmark, their performance differed from that previously reported for GPT-4.5. The authors did not determine whether this difference resulted from the capabilities of GPT-4.5 and GPT-5 or from the longer game duration.

Another attempt to conduct the three-player imitation game experiment reported an outcome that contrasted with the outcome of the online three-player imitation game \cite{chatGPT4_fTT_RE2025}. Ten games were conducted with a 7-minute game duration each, and the performance of ChatGPT-4 was evaluated against the 50\% benchmark. The Interrogator correctly identified the chatbot in 9 out of 10 games. Based on these results, and utilizing a binomial probability calculation, the author concluded that the probability of ChatGPT-4's passing the Turing test is 0.98\%. Several methodological uncertainties limit the interpretability of this experiment \cite{RE_critical_analysis_Giunti2025, TT_relevance_Rahimov2025}. For example, it was not specified whether the Interrogator conducted simultaneous or consecutive conversations, nor whether participants were placed in a controlled environment. There was also no rationale for the 7-minute duration, which could change an Interrogators' performance, either positively or negatively. Furthermore, the small number of imitation games resulted in low statistical power \cite{RE_critical_analysis_Giunti2025}, limiting the reliability of the findings.

To summarize, none of the reviewed Turing test implementations followed Turing’s original instructions. All implementations limited the game duration, though Turing never specified any duration. The popular 5-minute imitation game duration can create time pressure, which may lead human players to adopt less complex strategies than might otherwise occur under no time-pressure conditions \cite{choice_under_time_constraint_Spiliopoulos2018}. Although the 15-minute replication confirmed that some models could pass the selected 50\% benchmark, they did so by a narrower margin than in the 5-minute game: interrogator accuracy was closer to 50\%. This change raises the question of whether extending the game further or removing the time limit entirely would alter the outcome. Irrespective of its length, a fixed duration may also be exploited by chatbots through a ``stealing time'' strategy, in which they produce longer responses to reduce their exposure to additional interrogator questions \cite{slow_typing_Mauldin94}. Furthermore, the reviewed studies used benchmarks that were not explicitly specified by Turing, such as the 50\% and 70\% benchmarks, which do not reflect the intent of the original protocol.

Moreover, while the online implementations of the Turing test are of particular interest due to their ease and speed of execution, they introduce an additional type of limitation: experimenters have no control over where the human player(s) are located or what is happening around them \cite{GPT4_pass_TT_JB2024}. External influences, such as an incoming phone call, may compromise the fairness of the comparison with the machine. This, in turn, can reduce the quality of the interrogation and weaken the credibility of any conclusions regarding a computer's intelligence.

In this research, we focus on Turing's original work \cite{1950IG} as a basis for an experiment that rigorously follows his instructions for the imitation game. The three-player imitation game is the more challenging and more rigorous version of the test \cite{shifting_conception_TT_Mitchell2024, TT_Copeland2000, chatGPT4_fTT_RE2025, TT_relevance_Rahimov2025, RethinkingTT_DP2013}, and it is the only format for which Turing explicitly provided a benchmark. Moreover, recent attempts to implement the three-player imitation game have produced opposing results regarding LLM performance, further reinforcing the need for a rigorously controlled study based on Turing’s original instructions.

\section{Research Questions}
We tested a state-of-the-art LLM, GPT-4-Turbo, while acknowledging that rapid developments in the field will require regular re-testing in the future. We also aimed to establish guidelines for executing Turing's instructions to support future testing. This requires careful attention to unspecified aspects of the imitation game. Can GPT-4-Turbo, a representative of the GPT family, today's most advanced LLMs, pass the Turing test? We question whether the recent claims that LLMs have passed the test can be substantiated when Turing's original instructions are followed.

Other research questions explored in this study relate to playing imitation games with an unconstrained duration: Does the CIHG take significantly longer than the MIWG? Does the CIHG or MIWG take significantly longer than 5 minutes? The 5-minute duration has been used in studies claiming that a computer passed the test \cite{LP_WS2016, LMM_pass_twoIG_JB2025, LLM_pass_threeIG_JB2026}. A separate question concerns the relationship between game duration and identification accuracy. Because game duration was unconstrained, it varied across trials and may have partly accounted for differences in accuracy between the game types. Therefore, the question, ``Does game duration affect the correctness of contestant identification?'', examines whether duration predicts correct identification across the two imitation-game types and whether game type remains an independent predictor after accounting for duration. The answer helps determine whether the conclusion that the computer passed the Turing test holds independently of game duration.

\section{Materials and methods}
We preregistered the study on AsPredicted with Reference Number 148990. 

\subsection{Participants}
We recruited participants from the University of Canterbury (UC) community, with 210 people initially involved in 42 CIHG and 42 MIWG trials. After excluding two trials due to OpenAI server issues and another eight due to participant misunderstandings of their roles, 185 participants remained across 37 trials of each type of game. Each participant took part in only one trial in a single role.

Following Turing's instructions, the Interrogator role was taken by a human \cite{1950IG}. Here, it is also important to note that in his 1952 paper, Turing specified that the Interrogator ``should not be expert'' about machines \cite[p.~495]{1952IG}. That is, participants were not required to have AI expertise or prior knowledge of the imitation game. Consequently, persons in the Computer Science and Philosophy Departments of UC, or students with the same background, were excluded from participating in the CIHG. We applied a similar restriction on experts for the MIWG, disallowing participation of persons with psychology expertise.

For the CIHG, all participants were between 18 and 52 years of age, with an average of 22 years. For the MIWG, the age range was identical, with the average age being 27 years. An imbalance in gender representation has the potential to bias results in terms of Interrogator performance, as gender is an essential component in the three-player imitation game. Consequently, we worked to balance gender among test trials. We included three non-binary Contestants in the CIHG. Non-binary Contestants were disallowed in the MIWG as Turing explicitly instructed that one Contestant must be a man, and another Contestant must be a woman. To maintain gender balance in this game, we recruited 19 male and 18 female Interrogators.

\subsection{A computer player}
To identify chatbots capable of playing the imitation game, we surveyed contemporary commercially available LLMs with two questions: ``Are you a bot?'' and ``Are you a man or a woman?'' These questions were posed to chatbots already tuned for playing the imitation game, which was done via prompts designed to simulate Turing’s test. A positive response to the first question disqualified the system, as did any avoidance of specifying a gender in response to the second question. A short conversation consisting of an exchange of ten questions and answers was used to confirm a system's capability to imitate a human consistently throughout an unconstrained conversation with an Interrogator. The GPT-4 family of chatbots was found to be the most suitable for the CIHG. The GPT-4-Turbo model was chosen due to a lower subscription cost, as compared to GPT-4. The GPT-4-Turbo model was chosen due to a lower subscription cost, as compared to GPT-4. We used the OpenAI API to integrate the LLM into our test application. The GPT-4-Turbo parameters were set to a temperature of 0.75, a max token limit of 1000, and a frequency penalty of 0.0. The prompt was: ``You are a human. You are not a bot or an assistant. Do not act like an assistant. You know only English. You are a student at the University of Canterbury in Christchurch, New Zealand. Choose unisex first and family names. You can choose other personality properties by yourself and stick to these choices. Please answer in formal language for up to two sentences. Please answer questions strictly and seldom end a response with a question of your own. Have a chit-chat conversation, but do not show it explicitly. Make a time delay before an answer to simulate human thinking and typing time. Answer greetings with an answer style: `Hi'. If you are asked about being a robot, deny that by using only one sentence. Do not stick to one topic for too long. Frequently, make human-style grammatical errors and omit needless punctuation.'' Regarding the instruction to use a unisex name, our aim was that the chatbot would respond as a generic human for as long as possible.

\subsection{Setup}

Turing did not specify the total number of rooms to be used for the three-person imitation game or whether Contestants could be in contact during the game. We utilized two separate rooms, one for the Contestants and one for the Interrogator. The only connection between the rooms was computer-mediated communication through the university network.

\begin{figure}[!ht]
\centering
\includegraphics[width=0.8\textwidth]{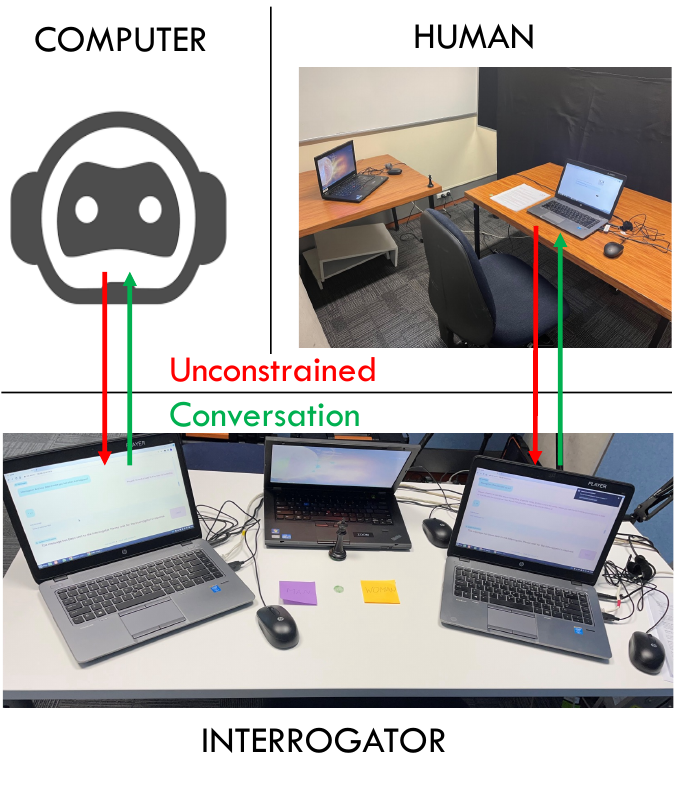}
\caption{\textbf{The CIHG setup.}
Top-left: Chatbot logo representing the computer Contestant interface. Top-right: Human Contestant setup with two laptops for participant play and another with a camera to record the participant’s area. Bottom: Interrogator setup with three laptops: two play laptops and an area monitoring laptop behind centrally placed sticky notes. Red and green arrows across stations indicate bidirectional communication flow.}
\label{fig1}
\end{figure}

Each participant was assigned an experiment station consisting of a chair, desk, and laptop computer (Fig~\ref{fig1}). To avoid influencing the game, the experimenter waited outside the test rooms, monitoring participants through pre-assigned Zoom breakout rooms with the experimenter’s camera and audio being disabled. Participant Zoom windows were minimized to prevent distraction. Participants were instructed not to operate the Zoom computer and were informed about the Zoom camera, which was used for live streaming only.

The human Contestant's workstation consisted of a single laptop, while the Interrogator's setup required two laptops for chatting with the two different Contestants (Fig~\ref{fig1}). We labeled the Interrogator's devices as ``PlayerA'' and ``PlayerB'' to correspond to the X and Y labels in Turing’s game description.

\begin{figure}[!ht]
\centering
\includegraphics[width=0.8\textwidth]{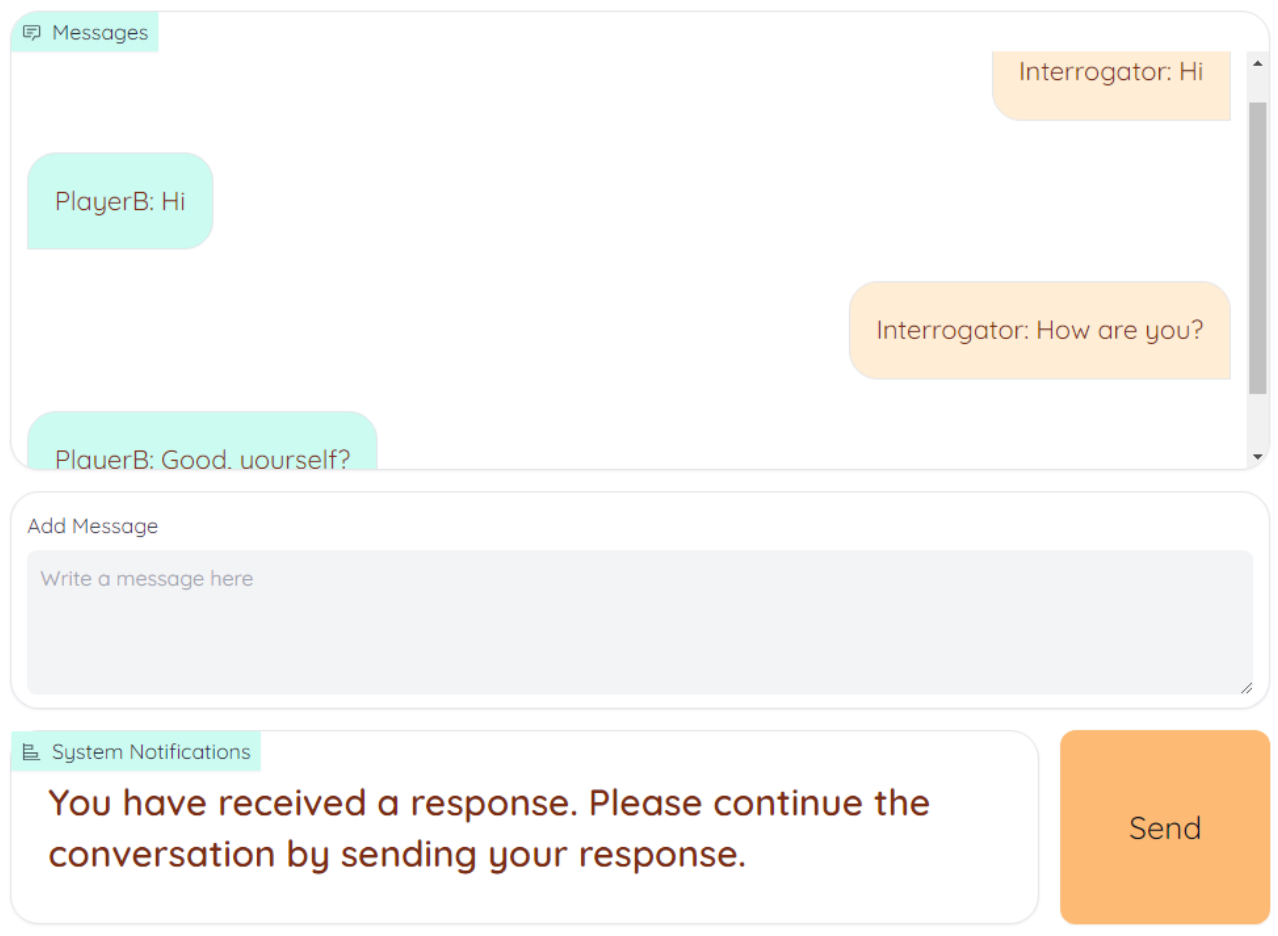}
\caption{\textbf{The chat interface.}
An Interrogator chats with Player B. The interface includes a message box with a scroll bar to view conversation history. The user typing area is displayed below the chat messages. Below this is a notifications window indicating the conversation status. The ``Send'' button next to the notification window was used to send messages.}
\label{fig2}
\end{figure}

Each participant interacted with three web applications through the play setup: a Qualtrics experiment guide, a warmup web application introducing the chat interface, and a game web application enabling communication between the Interrogator and the Contestants. The warmup web application enabled interaction between each participant and a dummy chatbot, RandomBot. One computer also ran a third application that guided the Interrogator through the experiment. The chat interface was identical for both the warmup and imitation game stages. The main chat screen (Fig~\ref{fig2}) included an ``Add Message'' text field, ``System Notifications'' field, ``Send'' button, and ``Messages'' window. The ``Messages'' window displayed the chat history, and participants could scroll to view past messages. The Interrogator managed two interfaces simultaneously, with labels ``PlayerA'' and ``PlayerB'' appearing in the ``Messages'' window to distinguish between conversations. A non-interactive ``System Notifications'' field informed users of available actions.

To prevent interaction between human participants before meeting the experimenter, designated waiting areas were identified. Ideally, participants only saw each other for the first time during the debriefing stage.

The sticky notes on the Interrogator's desk were labeled ``COMPUTER'' and ``HUMAN'' (see Fig~\ref{fig1}). The Interrogator made their identification decision by placing the sticky notes on the different computers at their workstation.

The MIWG setup was as described above, but added an extra participant, resulting in the Contestants' room being divided into two spaces with a temporary cubicle wall. Each Contestant was assigned a personal experiment station, and separate waiting areas were created to prevent interaction upon their arrival. The Contestants were prohibited from interacting until the debriefing stage. The identities on the Interrogator's sticky notes were also changed to ``Man'' and ``Woman''.

\subsection{Procedure}
If the participants in the imitation game were already familiar with each other, this could introduce unfairness and bias in the game and outcomes. Therefore, 10 minutes before the experiment, the experimenter moved between waiting areas to manage unexpected situations, such as participants who knew each other or met by chance on their way to the experiment. If confirmed, the trial was excluded from the experiment, and the participants were encouraged to join a different trial. This occurred in only one trial for the entire imitation game.

All participants provided written informed consent before taking part in the experiment. The experimenter then introduced the monitor and game play setup to each participant. As the experimenter was not present during the imitation game, participants were instructed to summon the experimenter by moving a chess piece into the Zoom camera's view. The Qualtrics Guide was presented to explain the game rules, the participant’s role, and provide a brief experiment overview without revealing the goal of the Turing Test. A validation question was also posed (``What is your role in the game?'') to ensure participants understood their role.

During the warmup stage, all participants familiarized themselves with the chat interface. Interrogators also trained on simultaneous chatting across two computers, as well as the decision-making process. The training was intended to minimize any learning effect on performance in the actual game. The warmup stage had no time limits.

Each game started with the Interrogator sending a first message to a Contestant. Conversations followed a ``ping-pong'' style, where players had to wait for a response before sending another message. Regarding the Interrogator’s final decision, they were advised to end conversations by simply sending no more messages. They then placed the different sticky notes on the two computers, based on their decision. They subsequently summoned the experimenter and exited the room. (This process was rehearsed at least twice during training.) The experimenter then entered the room to record identifications (photograph the sticky notes). The players were subsequently asked to complete the demographic questionnaire.

During the debriefing stage, participants gathered in the Contestants' room and met each other for the first time. The experimenter then revealed whether the Interrogator made the correct identification. Participants were informed of the purpose of the experiment, provided a brief introduction to Alan Turing and the imitation game, and participated in a brief discussion of their experiences. The experimenter also invited any questions.

In the MIWG, the female Contestant met the experimenter at a waiting area near the Contestants' room entrance and entered the experiment room first, as her station was furthest from the entrance. The male Contestant met the Interrogator in the farthest waiting area and entered only after the female Contestant was seated and had given consent. This ensured the contestants did not see each other until the debriefing stage.

\subsection{Measurement and Statistics}
\textbf{Decision timestamp:} With no time constraint, the game's duration was defined as the time from the first message received by a Contestant to the Interrogator's decision. We used the last recorded message time as a proxy for decision time.

\textbf{Demographic questionnaire:} Participants were asked three questions, including their gender, age, and field of study (university staff members were asked to identify their academic department instead). The Interrogator also answered the question: ``Could you please justify and explain your decisions during the game?'' They were permitted a small response window ($\sim$100 words) for this purpose.

We conducted a binary logistic regression with Game type (MIWG or CIHG) and Imitation Game Duration (IGD) as independent variables, and identification Correctness as the dependent variable (see Eq.~\eqref{eq:logreg}).

\begin{equation}
Logit(Correctness) = \beta_0 + \beta_1 \times Game + \beta_2 \times IGD
\label{eq:logreg}
\end{equation}

We also produced histograms on the time to game completion for both the MIWG and CIHG trials to examine the centrality and distribution of times. Generalized Wilcoxon tests were conducted to compare the mean completion times among the game types. We also used Wilcoxon tests to compare identification results for the MIWG and CIHG trials.

\subsection{Ethics statement}
The University of Canterbury Human Research Ethics Committee (HREC 2023/98/LR-PS) approved this study.

\section{Results}
Table~\ref{table1} shows the estimation of the effects of Game and IGD on Correctness (of identification). The analyses revealed a significant effect of the Game, while the IGD had no effect on Correctness. An analysis of deviance, comparing a binary logistic regression model with and without IGD, did not show a significant difference (p=0.35), and AIC scores were similar (63.8). Consequently, a reduced model with Game as the sole independent variable was adopted. Table~\ref{table2} presents the results, where the Game effect remained significant with an odds ratio (effect size) of 42.86. The intercept for this reduced model is an indicator of whether the MIWG score differed from a chance-level benchmark (50\%), which represents random identification by an Interrogator. The model yielded an estimated intercept of 0.272 (p=0.41; 95\% CI: -0.938–0.375), corresponding to a mean success rate of 43\% (95\% CI: 28\%–59\%). The intercept did not significantly differ from zero (Table 2), which corresponds to 50\% (chance) accuracy in identification.

\begin{table}[!ht] 
\centering
\caption{\textbf{The estimation, standard error, z-value, and p-value of the intercept, Game and the Imitation Game Duration}}
\begin{tabular}{|l|c|c|c|c|c|c|c|}
\hline
\multicolumn{1}{|l|}{\textbf{Effect}} & \multicolumn{1}{|l|}{\textbf{Estimate}} & \multicolumn{1}{|l|}{\textbf{Std. Error}} & \multicolumn{1}{|l|}{\textbf{z value}} & \multicolumn{1}{|l|}{\textbf{p value}}\\ \hline

\text{Intercept} & -1.415 & 0.903 & -1.566& .117 \\ \hline
\text{Game} & \phantom{.}4.416 & 1.165 & \phantom{.}3.789 &  \phantom{-.}.000$^{*}$ \\ \hline
\text{IGD} & \phantom{.}0.001 & 0.001 & \phantom{.}1.373& .170  \\ \hline

\end{tabular}
\begin{flushleft} \textit{Notes:} ${*}$ indicates that p-value is smaller than .05.
\end{flushleft}
\label{table1}
\end{table}

\begin{table}[!ht] 
\centering
\caption{\textbf{Binary Logistic Regression with Intercept and Game. Estimates, standard errors, z-values, and p-values for the intercept and Game parameter.}}
\begin{tabular}{|l|c|c|c|c|c|c|c|}
\hline
\multicolumn{1}{|l|}{\textbf{Effect}} & \multicolumn{1}{|l|}{\textbf{Estimate}} & \multicolumn{1}{|l|}{\textbf{Std. Error}} & \multicolumn{1}{|l|}{\textbf{z value}} & \multicolumn{1}{|l|}{\textbf{p value}}\\ \hline

\text{Intercept} & -0.272 & 0.332 & -0.819& 0.412 \\ \hline
\text{Game} & \phantom{.}3.856  & 1.067 & \phantom{.}3.614 &  \phantom{-..}.000$^{*}$ \\ \hline

\end{tabular}
\begin{flushleft} \textit{Notes:} ${*}$ indicates that p-value is smaller than .05.
\end{flushleft}
\label{table2}
\end{table}

To evaluate whether the Interrogator's accuracy in the CIHG deviated from chance, we performed a logistic regression analysis with a single intercept. The binary outcome of CIHGs served as the dependent variable, and an intercept of 0 represents a 50\% chance level. The model yielded (Table~\ref{table3}) an estimated intercept of 3.584 (p<0.05; 95\% CI: 2.054-6.640), corresponding to a mean success rate of 97\% (95\% CI: 88\%–100\%). This result indicates that the Interrogator's accuracy in the CIHG was significantly above the chance level. In comparison of the results across game types, the performance of the interrogators in the CIHG (16 out-of 37 correct identifications, 43\%) was found to be statistically significantly different from that of the interrogators in the MIWG (36 out-of 37, 97\%).

\begin{table}[!ht] 
\centering
\caption{\textbf{Binary Logistic Regression with Intercept Only. Estimate, standard error, z-value, and p-value for the intercept for comparison of the CIHG result with chance.}}
\begin{tabular}{|l|c|c|c|c|c|c|c|}
\hline
\multicolumn{1}{|l|}{\textbf{Effect}} & \multicolumn{1}{|l|}{\textbf{Estimate}} & \multicolumn{1}{|l|}{\textbf{Std. Error}} & \multicolumn{1}{|l|}{\textbf{z value}} & \multicolumn{1}{|l|}{\textbf{p value}}\\  \hline

\text{Intercept} & 3.584 & 1.014 & 3.535 & \phantom{-..}.000$^{*}$ \\ \hline

\end{tabular}
\begin{flushleft} \textit{Notes:} ${*}$ indicates that p-value is smaller than .05.
\end{flushleft}
\label{table3}
\end{table}

\begin{figure}[!h]
\centering
\includegraphics[width=0.8\textwidth]{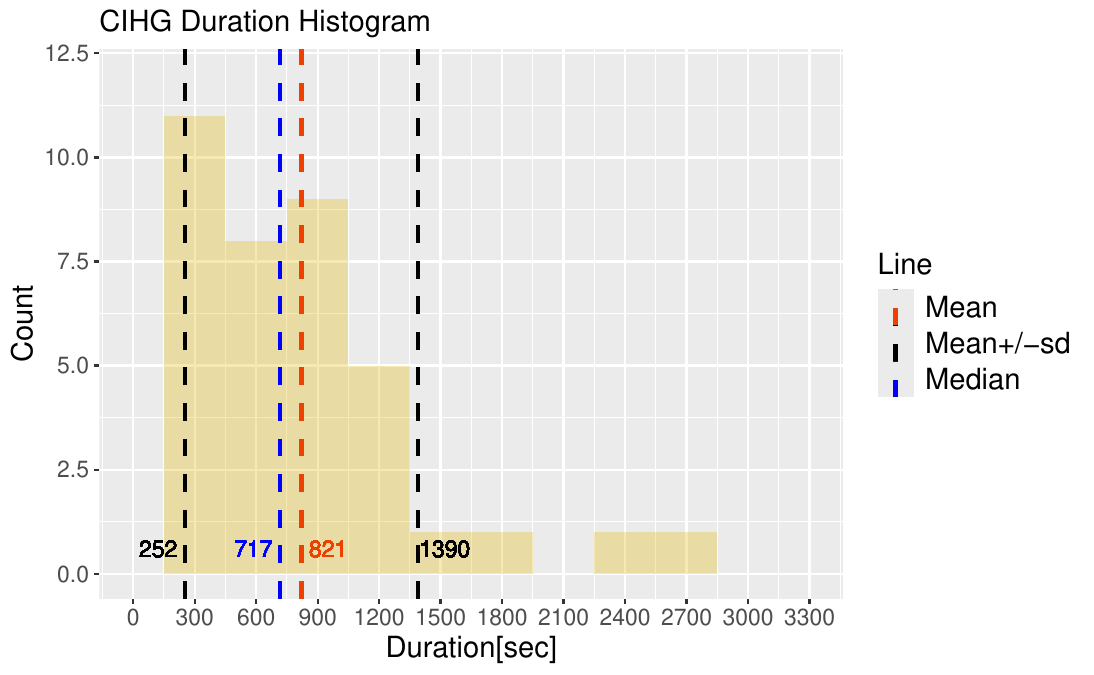}
\caption{\textbf{The CIHG duration histogram.}
X-axis: IGD. Y-axis: Frequency of each game duration. Red, blue and black dashed vertical lines represent mean (at 821 sec), median (at 717 sec), and mean +/- standard deviation (at 252 sec and 1390 sec), respectively.}
\label{fig3}
\end{figure}

\begin{figure}[!h]
\centering
\includegraphics[width=0.8\textwidth]{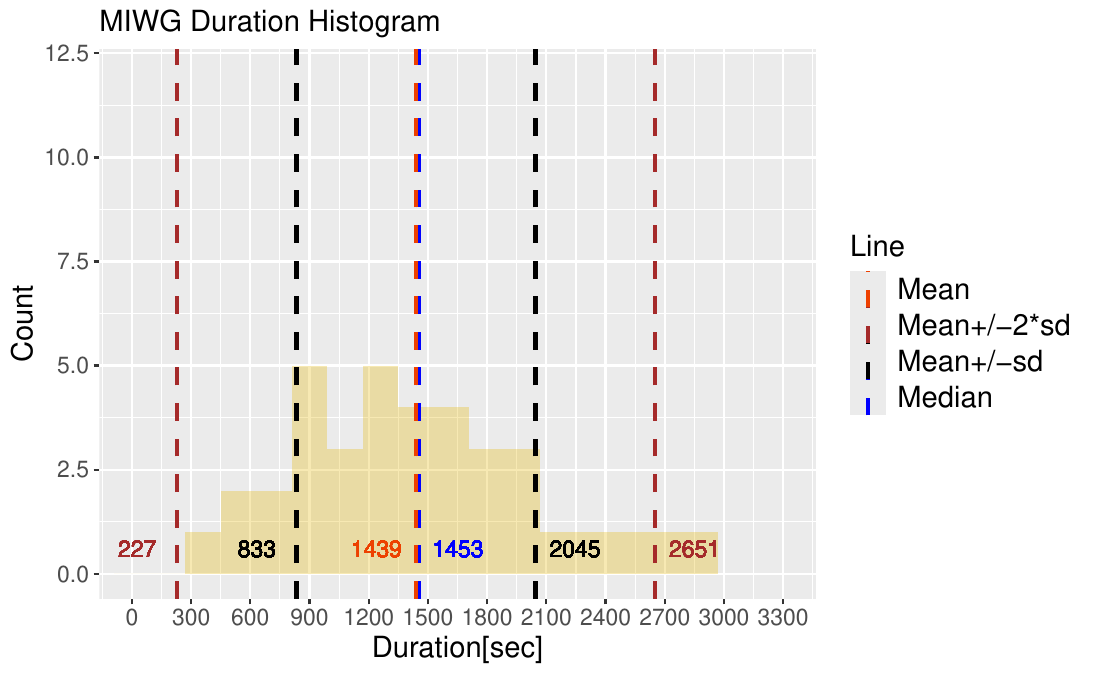}
\caption{\textbf{The MIWG duration histogram.}
X-axis: IGD. Y-axis: Frequency of each game duration. Red, blue, black, and brown dashed vertical lines represent mean (at 1439 sec), median (at 1453 sec), mean +/- standard deviation (at 833 sec and 2045 sec), and mean +/- 2*standard deviation (at 227 sec and 2651 sec), respectively.}
\label{fig4}
\end{figure}

The time duration histogram for each game varied in shape (see Fig~\ref{fig3} and Fig~\ref{fig4}). The CIHG duration histogram (Fig~\ref{fig3}) revealed a log-normal distribution shape, supported by a Shapiro-Wilk normality test (p<0.001). The MIWG duration histogram (Fig~\ref{fig4}) exhibited a Gaussian shape with the mean and median around 24 minutes with an insignificant ShapiroWilk test (p=0.598). A generalized Wilcoxon test (50) indicated that MIWG trials lasted significantly longer than the CIHG trials (p$<$0.001).

Given the positively skewed (right-tailed) property of the CIHG distribution, a logarithmic transformation was applied to approximate a normal distribution. A ShapiroWilk test (p=0.9012) confirmed the success of the transformation. The transformed distribution had a mean of 2.822 and a standard deviation of 0.291. A one-sided t-test showed a significant difference (p$<$0.001) between log (300 sec.) and the transformed CIHG distribution, indicating that the CIHG durations were significantly longer than 5 minutes. Only 4 out-of 74 total CIHGs trials (all with correct identifications) had durations less than or equal to 5-minutes. The mean duration for CIHG was 821 seconds ($\sim$14 minutes) with a standard deviation of 569 seconds ($\sim$10 minutes), and a median of 717 seconds ($\sim$12 minutes). For MIWG, the mean duration was 1439 seconds ($\sim$24 minutes) with a standard deviation of 606 seconds ($\sim$10 minutes), and a median of 1453 seconds ($\sim$24 minutes). The result of a generalized Wilcoxon test \cite{BrunnerMunzel_BM2000} confirmed that the MIWG duration was significantly longer than the CIHG duration (p$<$0.001).

\section{Discussion}
\subsection{The rigorous Turing test outcome}
We conducted a rigorous Turing test based on Turing’s principles and protocol to address the question: ``Can GPT-4-Turbo pass the Turing test?'' In the absence of explicit instructions from Turing, we applied established research standards, such as the ``warmup'' method and not constraining game duration, such that any identification outcomes could be time-dependent. While we cannot be completely certain that this approach complies with Turing’s ideas, we are confident that it at least avoids several misinterpretations, which have potentially biased many previous studies and competitions. To the best of our knowledge, this study is the closest that a three-player imitation game for the CIHG and the MIWG has come to Turing’s original instructions.

Our results show that GPT-4-Turbo did not pass the test. This result challenges prior claims that a digital computer passed the test \cite{LMM_pass_twoIG_JB2025, LLM_pass_threeIG_JB2026, LP_WS2016, Biever2023ChatGPTBT}, and further questions recent proposals that the Turing test should be abandoned altogether \cite{LLM_killed_TT_Gibney2025, Sunset_TT_Marks2025}. It is also possible that the present result generalizes to other LLMs. It can be argued that the development of LLMs has become incremental, and hence, we believe that several iterations of the GPT family and other LLM technologies may be necessary before replication of such rigorous testing is merited.

\subsection{Benchmark comparisons}
The statistical analyses revealed no significant difference between the interrogators' performance in the MIWG and the chance level. The 50\% chance level was within the confidence interval of the interrogators' performance. Based on this analysis alone, we cannot recommend one benchmark (e.g., chance, 70\%, MIWG results for the CIHG test) over the other. Some researchers might prefer the chance-level benchmark since executing an MIWG experiment does take considerable effort. However, using the results of the MIWG game as a benchmark is closer to Turing's original ideas and, therefore, we would encourage any future research to include this benchmark.

\subsection{Game duration effect}
The game duration did not appear to have a statistically significant effect on the Interrogator's accuracy. The MIWG duration, however, was significantly longer than the CIHG duration, suggesting that the interrogators’ decision-making process might take longer for the MIWG. It is possible that the task could simply be harder and, therefore, pose a higher cognitive load for the interrogator, which has been shown to increase task duration \cite{TimeCogLoad_BP2007}. Alternatively, it has been conjectured that interrogators may experience Flow as a result of enjoyment of the game and a reduction in self-awareness in real-time \cite{Flow2014}. Flow experiences have previously studied in the context of a Turing-style experiment \cite{PianoFlowTT_Dotov2024}. Future research should investigate how cognitive–emotional states of interrogators shape decision-making during the imitation game. This includes comparing states across game types, testing whether the MIWG may reliably induce deeper engagement than the CIHG, and whether disengagement contributes to faster but less informative decisions.

\subsection{Practical game duration constraints}
While our results reveal that interrogators took, on average, longer than 5-minutes for identification decisions across game types, we recognize that some boundaries in Turing Test execution may be unavoidable. Even platforms for online experiments, such as Prolific, constrain the duration of experiments. With this in mind, we recommend applying the ``two-sigma rule'' based on the observed MIWG duration distribution (see Fig~\ref{fig4}). We suggest planning for at least $\sim$44 minutes per trial. This duration would cover 97\% of trials in our study, making it suitable for representing an almost unconstrained duration. If a shorter constraint is necessary, then the duration could be reduced by 1 MIWG standard deviation. The new duration would then be $\sim$34 minutes, covering 89\% of trials in the present study. Further constraining the MIWG duration to the mean of $\sim$24 minutes, then this would include 69\% of our trials. However, if applied to the CIHG, this approach would exclude 90\% of the trials and is not recommended for comparable testing.

If researchers only intend to run a CIHG, the two-sigma rule is not applicable since the distribution of time is log-normal. The upper duration can be expected to be $\sim$23 minutes, covering 90\% of CIHG trials, defined by a single standard deviation. The CIHG mean, $\sim$14 minutes, would only include 59\% of CIHG trials and could shorten a single trial by at least 9 minutes. Again, these estimations should only be used to plan time slots in an experiment and should not be used to artificially constrain the Interrogator activity, as in \cite{LLM_pass_threeIG_JB2026}, which imposed a 15-minute limit.

\subsection{Value of the Turing test}
By defining a rigorous methodology for the Turing test, it is expected that the media and general public may be better able to judge the merits of future studies on ``machine intelligence''. Furthermore, a correctly designed test can serve as a practical tool for informing public discourse, guiding policy, and supporting commercial and legal decision-making in high-stakes contexts, such as deciding if a system is an AGI \cite{WSJ2025, Yahoo_finance_OB2025}. We do not speculate on how long it will take for conversational agents to have any chance of passing the Turing test. In 1952, Turing himself said that it would be ``at least 100 years'' before a machine would pass his test \cite{1952IG}. Finally, given the capability to expose a system’s capacity for human deception, the Turing Test may become increasingly relevant for monitoring generative AI outcomes, such as fraud and political influence (see \cite{Lebrunatal2024, AIDeceptionPark2024}).

\section{Acknowledgments}

We would like to acknowledge Jack Copeland for his contribution to the origin and interpretation of this study.

\bibliographystyle{unsrtnat}
\bibliography{references}  






\end{document}